\documentclass[twocolumn,prc,showpacs,preprintnumbers,
               amsmath,amssymb]{revtex4}
\usepackage{graphicx}% Include figure files
\usepackage{dcolumn}% Align table columns on decimal point
\usepackage{bm}% bold math
\usepackage{longtable}

\newcommand{\beq}{\begin{equation}}
\newcommand{\eeq}{\end{equation}}
\newcommand{\beqn}{\begin{eqnarray}}
\newcommand{\eeqn}{\end{eqnarray}}
\newcommand{\btab}{\begin{tabular}}
\newcommand{\etab}{\end{tabular}}

\usepackage{multirow,amsmath,booktabs}

\begin{document}

\title{Shell States of Neutron Rich Matter}

\author{C.~J.~Horowitz\footnote{e-mail:  horowit@indiana.edu} }
\affiliation{Nuclear Theory Center and Department of Physics,
Indiana University Bloomington, IN 47405}
\author{G.~Shen\footnote{e-mail:  gshen@indiana.edu}}
\affiliation{Nuclear Theory Center and Department of Physics,
Indiana University Bloomington, IN 47405}

\date{\today}

\begin{abstract}

The equation of state (EOS) for nuclear and neutron rich matter is investigated in a Relativistic Mean Field (RMF) model.  New shell states are found that minimize the free energy per baryon, calculated in a spherical Wigner-Seitz (WS) approximation, over a significant range of baryon densities.  These shell states, that have both inside and outside surfaces, minimize the Coulomb energy of large proton number configurations at the expense of a larger surface energy.   This is related to a possible depression in the central density of super heavy nuclei.  As the baryon density increases, we find the system changes from normal nuclei, to shell states, and then to uniform matter.

\end{abstract}


\maketitle

\section{\label{sec:introduction}Introduction}

% EOS for Supernovae and Neutron Stars

The equation of state (EOS) for dense nuclear matter plays
an important role in understanding the structure of neutron stars and
for simulations of core collapse supernovae.   Neutron stars probe the zero temperature EOS while supernovae involve the EOS at temperatures up to tens of MeV.
Below the neutron drip density, $\approx 2\times 10^{11}$ g/cm$^3$ where free neutrons appear, knowledge of the EOS can be extrapolated from the properties of laboratory atomic nuclei.  In contrast, current terrestrial experiments only constrain the
EOS above neutron drip density with rather big uncertainties.
Recently, there have been efforts from heavy ion collision experiments \cite{heavyions}, X-ray observations of isolated neutron stars \cite{x-rays}, and
theoretical calculations to understand the EOS beyond the neutron-drip point.

There have been many calculations of nonuniform matter at sub-nuclear density, below $3\times 10^{14}$ g/cm$^3$.  Negele and Vautherin, using a Skyrme Hartree-Fock calculation with two-body potentials \cite{Negele}, obtained the ground state within the spherical
Wigner-Seitz (WS) approximation.  In this approximation, the
unit cell of a crystal lattice is modeled as a sphere.  They found, as the system became more
neutron rich and the density increased, neutrons escaped and the system approached a uniform state near nuclear density.  Since then there have been many investigations \cite{more} with more sophisticated interactions and more complicated lattice configurations, such as cylinders or plates.  These non-spherical 'pasta' phases, which seem to appear
within a significant range of sub-nuclear densities are relevant for the structure of neutron star crusts \cite{crust1,crust2} and the dynamics of supernovae \cite{sn}.

% Existing results: Liquid drop model, Thomas-Fermi
For hot nuclear matter, assuming that the spherical WS cell
still remains a good approximation to the realistic lattice structure,
people have studied the EOS with different models. Using a phenomenological compressible
liquid-drop model, Lattimer and Swesty \cite{LSeos} produced an
equation of state for hot dense matter that has been widely used in supernova simulations.  Later, H. Shen et. al. \cite{HSeos} constructed an equation of state based on
Thomas-Fermi and variational approximations to a relativistic mean field energy functional.  Neither method takes into account the shell structure of finite nuclei or explores the full range of density distributions possible even in the spherical WS approximation.

% what is RMF and its relevance, comparison with potential model
In recent years, relativistic mean field models (RMF) have
provided a consistent description for the ground state properties
of finite nuclei, both along and far away from the valley of beta
stability \cite{HS1,HS2,SW,Rein,Ring}.  These models incorporate
the spin-orbit splitting naturally and the relativistic formalism
provides a framework to extrapolate  the properties of non-uniform
and uniform nuclear matter to high densities.  Furthermore there
is a close relation between phenomenalogical RMF models, that
simply fit parameters to properties of finite nuclei, and more
systematic effective field theory approaches that enumerate all of
the possible interactions that are allowed by symmetries.

In this work, we develop a RMF code to explore the equation of
state for non-uniform and uniform nuclear matter at finite
temperatures with a range of proton fractions and wide range of
baryon densities.  For non-uniform nuclear matter, we use the
spherical WS approximation to find the stable state for a charge
neutral system of neutrons, protons and electrons.  In this paper,
we will focus on the matter at a low temperature T =1 MeV.  It is
easier for us to find the ground state at T = 0 by extrapolating
the theory at a low temperature T $\neq$ 0 to T = 0, because the
occupations of nucleon levels vary more smoothly at finite
temperature.  In the future we will also study the EOS at higher
temperatures.

% shell states

For nonuniform matter we find new shell states which minimize the free
energy per baryon over a significant density range. Shell states
have inside and outside surfaces and they can minimize the Coulomb
energy of high $Z$ (large proton number) configurations at the expense of a larger surface
energy.   These shell states may be related to the appearance of a central depression in the density of super heavy nuclei \cite{heavy} because of their large coulomb energies.
The appearance of shell states may significantly change transport properties such as the shear viscosity and shear modulus of neutron rich matter.

% Structure of the paper

We compare the transition density, that we find in nonuniform RMF calculations, to that predicted from a stability analysis of uniform nuclear matter.  We start from uniform matter in equilibrium, and then find the critical density when the longitudinal collective mode, that describes density oscillations, becomes unstable.

The paper is organized as follows.  First in Section \ref{sec:formalism}, we give a brief
description of the RMF theory at finite temperature and explain the WS
approximation for non-uniform matter.  Next, Section \ref{analysis} presents the collective mode analysis for uniform matter in both the Vlasov formalism and using the RPA approximation.  In Section \ref{sec:results}, we present results for the EOS of nonuniform matter.  Finally in Section \ref{summary}, we summarize our results and comment on future work.

\section{\label{sec:formalism}Formalism}

% introduction to RMF

The RMF theory and its applications have been
reviewed in a number of places, see for example Refs.~\cite{SW,Rein,Ring}. The basic ansatz of the RMF theory is a Lagrangian density where nucleons interact via the exchange of sigma- ($\sigma$), omega- ($\omega$), and rho- ($\rho$) mesons,
and also photons ($A$).
\beqn\label{lagrangian}
    {\cal L}&=&\overline{\psi} [i{\gamma^\mu}
              {\partial_\mu}-m-{g_\sigma}\sigma - g_\omega
              \gamma^\mu\omega_\mu \nonumber\\
              &&-\
              g_\rho \gamma^\mu \vec{{\bf
              \tau}}\cdot \vec{\mbox{\boldmath$\rho$}}_\mu - e\gamma^\mu\frac
              {1+\tau_3}{2} A_\mu ]\psi\nonumber\\
            &&+\
            \frac{1}{2}\partial^\mu\sigma\partial_\mu\sigma-\frac{1}{2}m_\sigma^2\sigma^2-
              \frac{1}{3}g_2\sigma^3\ -\ \frac{1}{4}g_3\sigma^4\nonumber\\
            &&-\
              \frac{1}{4}\omega^{\mu\nu}\omega_{\mu\nu}+\frac{1}{2}m_\omega^2\omega^\mu\omega_\mu+
              \frac{1}{4}c_3\left(\omega^\mu\omega_\mu\right)^2\nonumber\\
            &&-\ \frac{1}{4}\vec{\mbox{\boldmath$\rho$}}^{\mu\nu}
              \cdot\vec{\mbox{\boldmath$\rho$}}_{\mu\nu}+
              \frac{1}{2}m_\rho^2\vec{\mbox{\boldmath$\rho$}}^\mu
              \cdot\vec{\mbox{\boldmath$\rho$}}_\mu -\
              \frac{1}{4}A^{\mu\nu}A_{\mu\nu}
\eeqn
Here the field tensors of the vector mesons and the
electromagnetic field take the following forms:
\beqn\label{tensor}
   \omega^{\mu\nu}&=&\partial^\mu\omega^\nu-\partial^\nu\omega^\mu,\nonumber\\
                     A^{\mu\nu}&=&\partial^\mu A^\nu -\partial^\nu A^\mu,\nonumber\\
                     \vec{\mbox{\boldmath$\rho$}}^{\mu\nu}&=&\partial^\mu\vec{\mbox{\boldmath$\rho$}}^\nu-
                     \partial^\nu\vec{\mbox{\boldmath$\rho$}}^\mu-g_\rho\vec{\mbox{\boldmath$\rho$}}^\mu\times
                     \vec{\mbox{\boldmath$\rho$}}^\nu\,.
\eeqn

In charge neutral nuclear matter composed of neutrons, $n$, protons, $p$, and electrons, $e$, there are equal numbers of electrons and protons.  Electrons can be treated as a uniform Fermi gas at high densities. They contribute to the Coulomb energy of
the $npe$ matter and serve as one source of the Coulomb potential.

The variational principle leads to the following equations of
motion \beqn\label{nucleon-motion}
     [\mathbf{\sl{\alpha}}\cdot\mathbf{\sl{p}}\ +\ V(\mathbf{r})\ +\
     \sl{\beta} (m\ +\ S(\mathbf{r}))] \psi_i\ =\ \varepsilon_i\psi_i
\eeqn for the nucleon spinors, where \beqn \label{Dirac}
     \left \{
     \begin{array}
        {l} V(\mathbf{r}) =\ \beta \{g_{\omega}\rlap{/}{\omega}_{\mu}\
        +\ {g_{\rho}} \vec{\tau}\cdot
        \vec{\rlap{/}\rho_{\mu}}\  +\ e \frac{ (1\ +\ \tau_3)}{2}\ \rlap{/}A_{\mu} \}\\
        S(\mathbf{r}) =\ g_{\sigma}\sigma\
     \end{array}
     \right.
\eeqn

and \beqn\label{K-G}
    \left \{
    \begin{array}{l}
       (m_{\sigma}^2\ -\ \nabla^2)\
       \sigma =\ -g_\sigma \rho_s -\ g_2 \sigma^2\ -\ g_3\sigma^3 \\
       (m_\omega^2\ -\ \nabla^2)\omega^\mu =\ g_\omega j^\mu -c_3\omega^\mu(\omega^\nu\omega_\nu)\\
       (m_{\rho}^2\ -\ \nabla^2) \vec{\rho}^{\mu} =\ {g_{\rho}}
       \vec{j}^\mu\\
       \label{speqa}\hspace{2.5em} -\ {\nabla}^2A^\mu =\
       e(j^\mu_{p}-j^\mu_e)
    \end{array}
    \right.
\eeqn for the mesons and photons, where the electrons are included
as a source of Coulomb potential.  The nucleon spinors provide the
relevant source terms: \beqn\label{source}
    \left \{
    \begin{array}{l}
      \rho_s =\ \sum_i\overline{\psi}_i\psi_i n_i \\
      j^\mu =\ \sum_i\overline{\psi}_i \gamma^\mu \psi_i n_i\nonumber\\
      \vec{j}^\mu =\ \sum_i \overline{\psi}_i \gamma^\mu
      \vec{\tau} \psi_i n_i\\
      \label{speqb} j^\mu_{p} =\ \sum_i\overline{\psi}_i\gamma^\mu\frac
      {1+\tau_3}{2}\psi_i n_i.
    \end{array}
    \right.
\eeqn At zero temperature, the summations run over the valence
nucleons only, since $n_i=\Theta(\varepsilon_F-\varepsilon_i)$
where $\varepsilon_F$ is the Fermi energy of the nucleons.  At
finite temperature, Fermi-Dirac statistics prescribes the
occupations of protons and neutrons as follows: \beq\label{occu}
    n_i\ =\ \frac 1 {e^{\beta(\varepsilon_i-\mu)}+1},
\eeq where $\mu$ is the chemical potential for neutron (proton).
In practice, we include all levels with $n_i>10^{-8}$.

Since the systems under consideration have temperatures of, at
most, tens of MeV, we neglect the contribution of negative energy
states, {\it i.e.}, the so-called no sea approximation. In a
spherical nucleus, there are no currents in the nucleus and the
spatial vector components of $\omega_\mu$, $\vec{\rho}_\mu$ and
$A_\mu$ vanish. One is left with the timelike components. Charge
conservation guarantees that only the 3-component of the isovector
$\rho_{0,3}$ survives. The above non-linear equations are solved
by iteration within the context of the mean field approximation
whereby the meson field operators are replaced by their
expectation values.

\subsection{Nonuniform Matter with Wigner-Seitz approximation}

The spherical Wigner-Seitz (WS) approximation is used to describe
non-uniform matter.  In general the unit cell of a crystal lattice
is a complex close-packed polyhedron.  This is approximated by a
spherical cell of the same volume.  We apply boundary conditions
on the wave functions at the edge of WS unit cell.  To achieve a
uniform density distribution for a free neutron gas, we require
that at the cell radius, all wave functions of even parity vanish,
and the radial derivative of odd parity wave functions also
vanishes \cite{Negele}.

% Lattice Coulomb Energy

In the WS approximation, the lattice Coulomb energy consists of
contributions from neighboring unit cells. This correction to the WS
Coulomb energy is important for determining the stable
configuration of WS cells and the transition density to uniform
matter, especially when the system has a large proton fraction.
We include the exact Coulomb energy in calculating the
free energy of WS cell of radius $R_c$.  Following the treatment in
Ref.~\cite{Kittel,Oyamatsu}, we calculate the Coulomb energy per
unit cell as, \beq W_c = \frac 1 2 \sum_{\vec{G}} ' \frac{I_{hkl}^2}{a^3
\vec{G}^2}, \eeq where $a$ is lattice constant defined by
$a^3=V_{cell}=\frac{4\pi} 3 R_c^3$, and \beq \vec{G} =
h\vec{A}+k\vec{B}+l\vec{C}. \eeq Here $h, k$, and $l$ are
integers, and $\vec{A}$, $\vec{B}$, and $\vec{C}$ are the
primitive transformation vectors of the reciprocal lattice. The
prime on the sum means that the point $\vec{G}$ = 0 is excluded.
The form factor $I_{hkl}$ is given by \beq I_{hkl} = \int_{cell}
\rho_p(\vec{r}) e^{-i\vec{G}\cdot\vec{r}} d\vec{r}. \eeq Oyamatsu
et al. \cite{Oyamatsu} assumed a uniform proton density inside the nucleus and found that the stable configuration is a Body-Centered Cubic (BCC) lattice.  Using realistic proton density distribution, we also find that a BCC lattice gives the lowest Coulomb energy. In principle, one needs to worry about the screening of Coulomb interactions by
electrons.  It turns out that screening contributions to the Coulomb energy are very
small \cite{crust1} and we will neglect them in our calculations.

With the approximations specified above, it is convenient to
perform a self-consistent relativistic mean field Hartree
calculation for nuclear wave functions inside a WS cell of radius
$R_c$, for given average baryon density $\rho_B$, proton fraction
$Y_p$ and temperature $T$.  The nucleon number inside the WS cell
is therefore $A = 4\pi R_c^3\rho_B/3$ and proton number $Z = Y_p
A$. The total energy of the WS cell, including the exact lattice
Coulomb energy, is obtained as follows, \beqn\label{energy} E_b
&=& E_{nucleon} + E_\sigma + E_\rho + E_\omega + W_c - mA \cr &=&
\sum_i \epsilon_i n_i -\ e \int \rho_p A_0(r) d^3r \ -\ \frac 1 2
\int d^3r \{g_\sigma \sigma \rho_s(r)\ \cr && +\ \frac 1 3
g_2\sigma^3\ +\ \frac 1 2 g_3\sigma^4\}\ -\ \frac 1 2 \int
d^3rg_\rho\rho_{0,3} j_{0,3}(r)\ \cr && -\ \frac 1 2 \int
d^3r\{g_\omega \omega_0\ j_0(r) -\ \frac 1 2 c_3\omega_0^4\} +\
W_c- mA. \cr && \eeqn The nucleon contribution to the entropy is
given by the usual formula, \beq\label{entropy} S_b = -k_B \sum_i
\left[ n_i \mathrm{ln} (n_i)\ + (1-n_i) \mathrm{ln}
(1-n_i)\right], \eeq where $n_i$ is given in Eq.~(\ref{occu}).
With Eqs.~(\ref{energy}) and (\ref{entropy}), it is easy to obtain
the nucleon contribution to the free energy per baryon $f_b$,
\beq\label{freeb} f_b = F_b/A = (E_b\ -\ T S_b)/A. \eeq We also
include the contribution of free electrons.  Explicit formulas for
the electron contribution to the free energy per baryon $f_e$, and
entropy $S_e$ of the electron gas are presented in
Ref.~\cite{LSeos}.  Finally, the complete free energy per baryon
is \beq\label{free} f = f_b + f_e. \eeq

\subsection{Stability analysis}\label{analysis}
\subsubsection{Semiclassical collective modes analysis}

% give the formalism and how we do the calculation

The unstable collective modes of uniform $npe$ matter at finite
temperature can be investigated in the RMF by using a Vlasov formalism
\cite{Brito06}. Here we give a short summary to make this paper more self-contained.

The distributions of particles (antiparticles) denoted as + (-),
at position $\mathbf{r}$, time $t$, and momentum $\mathbf{k}$
($\mathbf{-k}$) are described via the phase space distribution
functions as following: \beq f(\mathbf{r},\mathbf{k},t)_{\pm}\ =\
\mathrm{diag}(f_{p\pm},f_{n\pm},f_{e\pm}). \eeq The time evolution
of distribution functions are determined by the Vlasov equation:
\beq \frac{\partial f_{i\pm}}{\partial t}\ +\
\{f_{i\pm},h_{i\pm}\} = 0, i = p, n, e, \eeq where \{\ ,\ \}
denotes the Poisson brackets. $h_{i\pm}$ is the one body
Hamiltonian, which can be derived from Eq.~(\ref{lagrangian}) for
neutrons and protons.  For electrons, $h$ is the normal QED
Hamiltonian.  Small deviations $\delta f_{\pm}$ of the
distribution functions around the equilibrium state can be
calculated with generating functions \beq
S(\mathbf{r},\mathbf{k},t)_{\pm}\ =\
\mathrm{diag}(S_{p\pm},S_{n\pm},S_{e\pm}), \eeq such that \beq
\delta f_{\pm}\ =\ \{S_{\pm},f_{0\pm}\}\ =\
\{S_{\pm},k^2\}\frac{df_{0\pm}}{dk^2}, \eeq where $f_{0\pm}$ are
equilibrium distribution functions.  In terms of the generating
functions, one can obtain linearized Vlasov equation for $\delta
f_{\pm}$.  From the equations of motion for meson and photon
fields, one can also get the linearized equations for the
oscillation fields.

The dominant collective modes at low temperatures are the
longitudinal modes (with momentum $\mathbf{p}$ and frequency
$\Omega$), which can be described by the ansatz \beq \left(
\begin{array} {c} S_{j\pm}(\mathbf{r},\mathbf{k},t) \\
                  \delta\sigma \\
                  \delta B^\mu
                  \end{array} \right)\ =\ \left( \begin{array}{c} S^j_{\Omega\pm}(k,\cos\theta) \\
                  \delta\sigma_\Omega \\
                  \delta B^\mu_\Omega
                  \end{array} \right)\ e^{i(\Omega
                  t-\mathbf{p}\cdot\mathbf{r})}, \eeq
where $j = p, n, e$, $B^\mu = \omega^\mu, \rho^\mu, A^\mu$
represents the vector fields and photon fields. $\theta$ is the
angle between $\mathbf{p}$ and $\mathbf{k}$. After transforming
the unknown variables to the density oscillations of $p, n, e$, we
can obtain the following matrix equation for the eigenmodes
\beq\label{dispersion} D(\Omega) \left(
\begin{array}{c}
\delta\rho^S_{\Omega p} \\
\delta\rho^S_{\Omega n} \\
\delta\rho_{\Omega p} \\
\delta\rho_{\Omega n} \\
\delta\rho_{\Omega e} \end{array} \right)\ =\ 0, \eeq where
$\delta\rho^S_{\Omega p}$ and  $\delta\rho^S_{\Omega n}$ are the amplitudes of the oscillating scalar densities of protons and neutrons, respectively,
and $\delta\rho_{\Omega p}$, $\delta\rho_{\Omega n}$, and
$\delta\rho_{\Omega e}$ are the amplitudes of the oscillating proton, neutron, and
electron densities.  The entries of the matrix $D(\Omega)$ are
given in Ref.~\cite{Brito06}.

The spectrum of collective modes, or the dispersion relation $\Omega=\Omega(p)$, is
determined by the eigen condition $D(\Omega)=0$. The instability of
collective modes can be deduced directly from the static limit
of the dispersion relation, {\it i.e.}, if $D(\Omega = 0) \leq 0$, the
collective modes will be unstable and exponentially growing. This
indicates a phase transition to non-uniform matter.

\subsubsection{RPA at finite temperature}

We now perform a second very similar stability analysis using the random phase approximation (RPA).  This provides an independent check of the Vlasov approach.   A detailed account of the finite temperature RPA method for
Quantum Hadrodynamics can be found in Ref.~\cite{ftRPA}.  Again we
present some information to demonstrate the idea.  The finite temperature Feynman rules yield the $(s,s')$ thermal component of the causal polarization insertion for particle
species $i$ $(i=n, p, e)$ as \beqn
\underline{\Pi}^{C(s,s')}_{\mu\nu(i)}\ &=&\ -i
\int\frac{d^4k}{(2\pi)^4}(-1)^{(s+s')}
\mathrm{Tr}[G_{H(i)}^{(s,s')}(k)J_\mu(i)\cr && \times
G_{H(i)}^{(s',s)}(k+q)J_\nu(i)], \eeqn where $G_{H(i)}^{(s,s')}$
is the $(s,s')$ thermal component of Hartree propagators of
species $i$. For $\gamma$(photon), $\omega$, and $\rho$ meson,
$J_\mu = \gamma_\mu$, and the polarization is a tensor
$\underline{\Pi}_{\mu\nu}$. For the $\sigma$ meson $J_\mu = 1$, and the
polarization is a scalar $\underline{\Pi}_s$. If $J_\mu =
\gamma_\mu$ and $J_\nu = 1$ (or vice versa), the polarization is mixed
 $\underline{\Pi}_{M\mu}$.

At finite temperature, Dyson's equation for the proper
longitudinal causal polarization is a matrix with
particle, Lorentz, and thermal indices. It can be written as \beq \underline{\widetilde{\Pi}}^c_L\ =\
\underline{\Pi}^c_L\ +\ \underline{\widetilde{\Pi}}^c_L
\underline{D}^0 \underline{\Pi}^c_L, \eeq where $\underline{D}^0$
is the lowest order thermal matrix for the meson/photon propagator. For
example, for scalar meson propagator is \beq \underline{D}^0(k) = \underline{M}_b \left(%
\begin{array}{cc}
  \Delta_{F}(k) & 0 \\
  0         & -\Delta_{F}^+(k) \\
\end{array}%
\right) \underline{M}_b, \eeq where $\Delta_{F}(k)$ is the normal
Feynman propagator and $\underline{M}_b$ is the thermal matrix
defined as
\beqn \underline{M}_b =\left(%
\begin{array}{cc}
  \cosh\theta & -\sinh\theta \\
  \sinh\theta & \cosh\theta \\
\end{array}%
\right). \eeqn The angle $\theta$ is given by $\sinh\theta =
[\exp\beta|k_0|-1]^{-1/2}$, with inverse temperature $\beta$. The
thermal vector meson propagator has the same structure.

At finite temperature we can also define
\beq\underline{\varepsilon}_L =\ 1\ -\ \underline{D}^0
\underline{\Pi}^c_L, \eeq which allows us to derive the dielectric
function by the real part of its (11) component, \beqn
\varepsilon_L(q_\mu)\ =\ \mathrm{det}
\mathrm{Re}(\underline{\varepsilon}_L(q_\mu)^{11}) &=&\
\mathrm{det} \{1-\Delta_{F}\mathrm{Re}\Pi_L^{c(11)}\},\cr && \eeqn
where we have used
$\mathrm{Re}\Pi_L^{c(11)}=\mathrm{Re}\Pi_L^{c}$. The poles of the
dielectric function define the collective mode excitations of
uniform matter.  The dielectric function in the static limit determines
the stability of the modes. When \beq\label{ftcritic}
\varepsilon_L(q_0=0, q) \leq 0, \eeq the system is unstable against
small amplitude longitudinal collective oscillations (density
fluctuations).  This indicates a phase transition to non-uniform
matter.

The above equation is for one particle species and one boson interaction. This can be generalized to several particle species with
multiple interactions. For a system consisting of neutrons,
protons, and electrons, the interaction matrix and polarization
matrix are given as
\beqn D_L^0\ &=&\ \left(%
\begin{array}{cccc}
  D_g^0 & 0 & -D_g^0 & 0 \\
  0 & -D_s^0 & 0 & 0 \\
  -D_g^0 & 0 & D_g^0+D_v^0+D_\rho^0 & D_v^0-D_\rho^0 \\
  0 & 0 & D_v^0-D_\rho^0 & D_v^0+D_\rho^0 \\
\end{array}%
\right), \cr && \\ \mathrm{Re}\Pi_L^{c(11)}\ &=&\ \left(%
\begin{array}{cccc}
  \Pi_{00}^{e(11)} & 0 & 0 & 0 \\
  0 & \Pi_s^{n(11)}+\Pi_s^{p(11)} & \Pi_M^{p(11)} & \Pi_M^{n(11)} \\
  0 & \Pi_M^{p(11)} & \Pi_{00}^{p(11)} & 0 \\
  0 & \Pi_M^{n(11)} & 0 & \Pi_{00}^{n(11)} \\
\end{array}%
\right). \cr && \eeqn The explicit formula for these polarizations
are listed in Ref.~\cite{ftRPA}. For the Lagrangian in
Eq.~(\ref{lagrangian}), the corresponding meson propagators are
\beqn D_g^0\ &=&\ \frac{4\pi\alpha}{q^2}, \cr D_s^0\ &=&\
\frac{g_\sigma^2}{q^2+m_\sigma^{*2}}, \cr D_v^0\ &=&\
\frac{g_\omega^2}{q^2+m_\omega^{*2}}, \cr D_\rho^0\ &=&\
\frac{g_\rho^2}{q^2+m_\rho^2}, \eeqn where the effective masses
are defined as \beqn m_\sigma^{*2}\ &=&\ m_\sigma^2\ +\
2g_2\sigma\ +\ 3g_3\sigma^2, \cr m_\omega^{*2}\ &=&\ m_\omega^2 +
3c_3\omega_0^2. \eeqn

\section{\label{sec:results}Results}

%\subsection{WS approximation: phase transitions at T = 1 MeV}
% sample: T = 1 MeV

% Yp = 0.4, F vs Rc
In this work we use the NL3 effective interaction \cite{NL3} that
has been successful in reproducing ground state properties of
stable nuclei and the saturation properties of symmetric nuclear
matter. The values of parameters for the NL3 effective interaction
are listed in Table \ref{tab:para}. In this section, we present
results in the spherical WS approximation for non-uniform matter
at a temperature of $T = 1$ MeV.

To determine the minimum free energy per baryon $f$ in
Eq.~(\ref{free}) at a specified baryon density $\rho_B$, one must
search over the cell radius $R_c$.  When $R_c$ is large and
$\rho_B$ is beyond the neutron drip density, one will need to take
into account a large number of levels, since the nucleon number is
$A = 4\pi R_c^3\rho_B/3$.  The fact that the matter is at finite
temperature will require one to consider even more levels.  For
example, when $\rho_B$ = 0.080 fm$^{-3}$, $Y_p$ = 0.4, and $R_c$ =
23.5 fm, there are 4892 nucleons inside one WS cell and one needs
to include 427 (419) neutron (proton) levels. It is hard to
achieve self-consistency for the mean fields with a large number
of levels.  To ensure the convergence of the self-consistent
iterations for the mean fields, it is important to have a good
initial guess for the mean field potentials.  Here our strategy is
to employ the convergent potentials for a nucleus with a smaller
$R_c$ as the starting guess for a nucleus with a slightly larger
$R_c$. In this way, new shell states with large cell radii are
found.  These can minimize the free energy over a significant
density range.

\begin{table}[h]
\centering \caption{NL3 effective interaction. The nucleon masses
are $M$ = 939~MeV for both protons and neutrons and $c_3=0$. } \label{tab:para}\btab{cccccccc}
\hline \hline $g_\sigma$ & $g_\omega$ & $g_\rho$ &
$g_2$ & $g_{3}$  & $m_\sigma$ &
$m_\omega$ & $m_\rho$\\
 & & & (fm$^{-1}$)  & & (MeV) & (MeV) & (MeV) \\
\hline
10.217&12.868&4.474&-10.431&-28.885& 508.194&782.5& 763\\
\hline \etab
\end{table}

\begin{figure}[htbp]
 \centering
 \includegraphics[height=8cm,angle=-90]{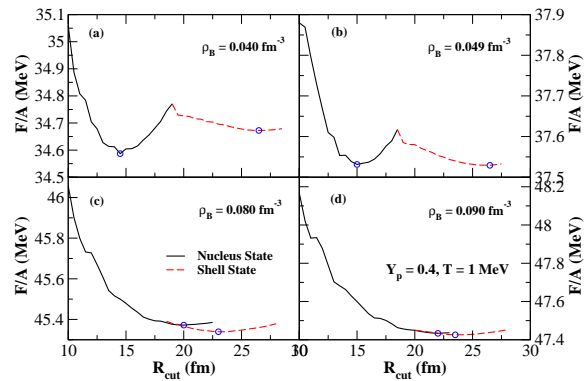}
 \caption{(Color online) Free energy per baryon versus Wigner-Seitz cell
 radius $R_c$ for baryon densities of $\rho_B$ = 0.040 (a),
0.049 (b), 0.080 (c), and 0.090 fm$^{-3}$ (d). The proton fraction
is $Y_p = 0.4$ and the temperature is $T$ = 1 MeV. The two circles
denote minima for the nucleus (small $R_c$) and shell states
(larger $R_c$).}
 \label{Yp0.4_mini}
\end{figure}

\begin{figure}[htbp]
 \centering
 \includegraphics[height=8cm,angle=-90]{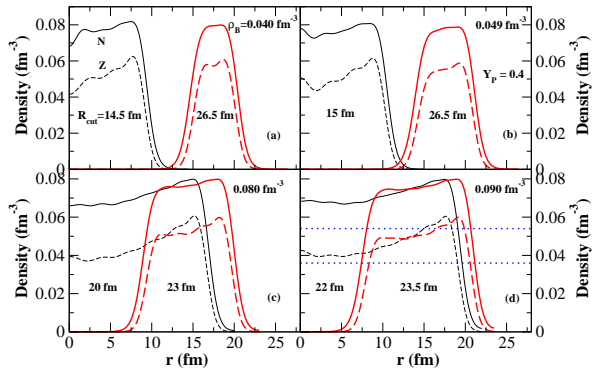}
 \caption{(Color online) The neutron (solid line) and
proton (dash line) density distributions for the Wigner-Seitz
cells with baryon densities of $\rho_B$ = 0.040 (a), 0.049 (b),
0.080 (c), and 0.090 fm$^{-3}$ (d). The proton fraction is $Y_p =
0.4$ and $T$ = 1 MeV. The red curves give density distributions
for shell states. The black curves give density distributions for
nucleus states. The corresponding cell radii are shown in each
panel. The dotted lines in the lower right panel are neutron
(upper) and proton (lower) densities for uniform matter.}
 \label{Yp0.4_density}
\end{figure}

We show in Fig.~\ref{Yp0.4_mini} an example of how shell states were found.  Here the free energy per baryon versus WS cell radius is displayed for baryon densities of 0.040, 0.049, 0.080, and 0.090 fm$^{-3}$ when $Y_p$ =0.4.  For each baryon density, there are two minima of free energy per baryon at different WS radii, denoted as open circles. The
neutron and proton density distributions of the WS cells with these
two radii are shown in Fig.~\ref{Yp0.4_density}. The first minimum
at smaller WS cell radius corresponds to a nucleus with a normal
density distribution. The second minimum at larger WS cell radius
corresponds to a shell shaped density distribution.  Here the nucleus has both outside and inside surfaces and the neutron and proton densities are only non-vanishing at intermediate $r$.  We call the first minimum a nucleus state and the second minimum a shell
state.  For the shell state, the sum of neutron and proton
densities (at intermediate $r$) is close to the saturation
density of nuclear matter $\sim$ 0.15 fm$^{-3}$, as one expects from nuclear saturation.

At $\rho_B$ = 0.049 fm$^{-3}$, the two minima of free energy per
baryon become degenerate, as shown in upper right panel of
Fig.~\ref{Yp0.4_mini}.  Below 0.049 fm$^{-3}$, the first minimum
with smaller cell radius corresponding to a normal nucleus is the
absolute minimum and therefore the equilibrium state ({\it e.g.},
panel a of Fig.~\ref{Yp0.4_mini}). On the other hand, above 0.049
fm$^{-3}$, the second minimum corresponding to a shell state is
the true equilibrium state ({\it e.g.}, panels c and d of
Fig.~\ref{Yp0.4_mini}). Therefore 0.049 fm$^{-3}$ is the critical
baryon density when the density distribution inside the WS cell
changes from a normal nucleus to a shell state.  In the lower
right panel of Fig.~\ref{Yp0.4_density}, we also show the uniform
neutron and proton density distributions by dotted lines when
$\rho_B$ = 0.090 fm$^{-3}$.  As will be shown later, 0.090
fm$^{-3}$ is the transition density from a shell state to uniform
matter.

%Yp = 0.4: F vs Rc; density distributions.

% shell states: interpretations

There is an intuitive interpretation of shell states by
considering the competition between the surface and Coulomb
energies.  When the Coulomb energy is more than half of the
surface energy, the nucleus can decrease its Coulomb energy, via
increasing its average radius and surface area, so that it
minimizes the total energy.  One possibility is to push the
nucleons out spherically and form a shell shape.  Note, this
change is possible even if one assumes spherical symmetry.   The
nucleus could also deform its shape to a non-spherical
configuration to minimize the total energy. It may be that, for
very neutron rich matter in certain regions of density, the most
stable configuration is non-spherical \cite{Gogelein}. Therefore
shell states may be related to the appearance of complex pasta
phases \cite{pasta, crust1, Gogelein}. Perhaps one can think of
shell states as ``spherical pasta".

\begin{figure}[htbp]
 \centering
 \includegraphics[height=6cm,angle=-90]{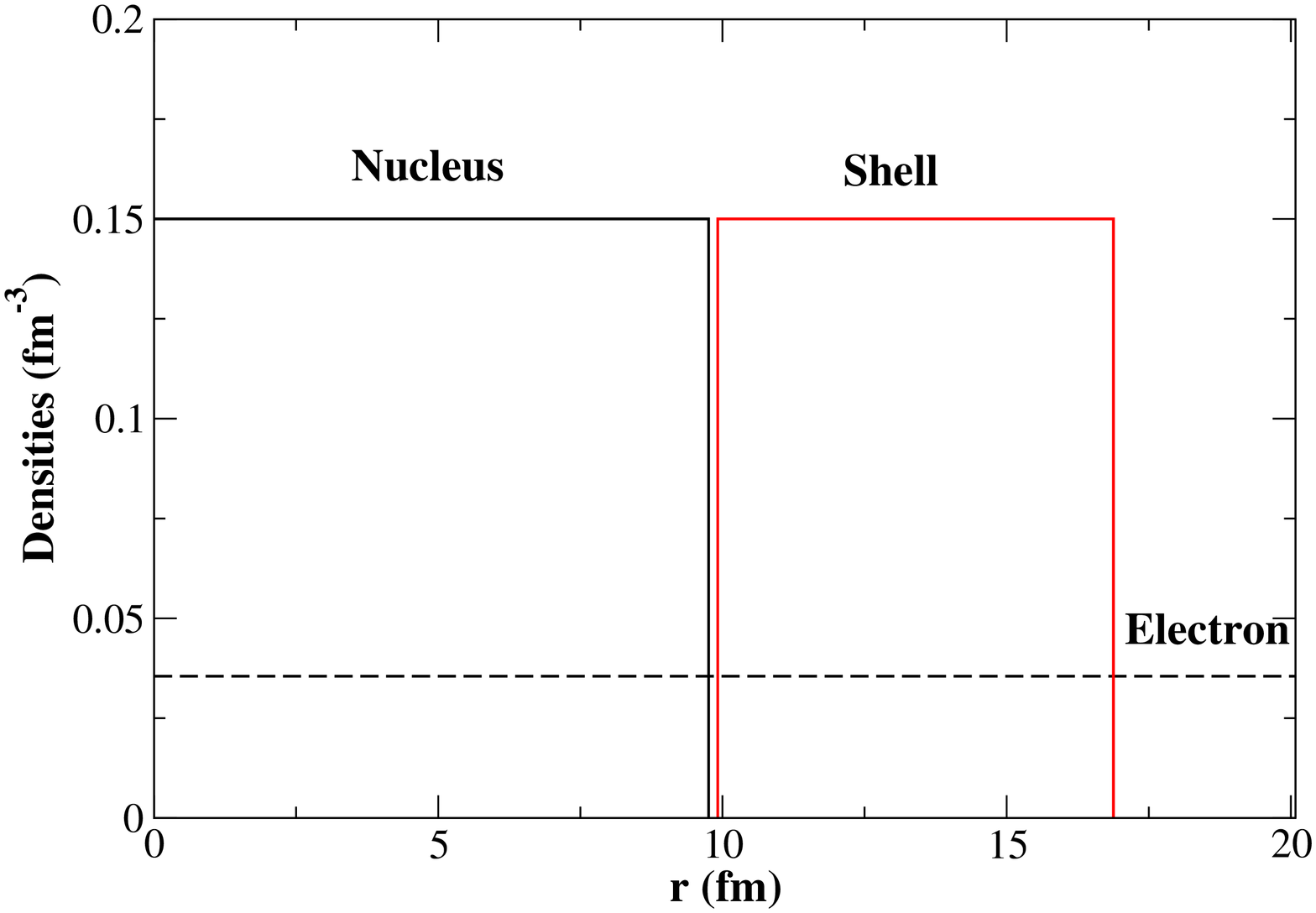}
 \includegraphics[height=6cm,angle=-90]{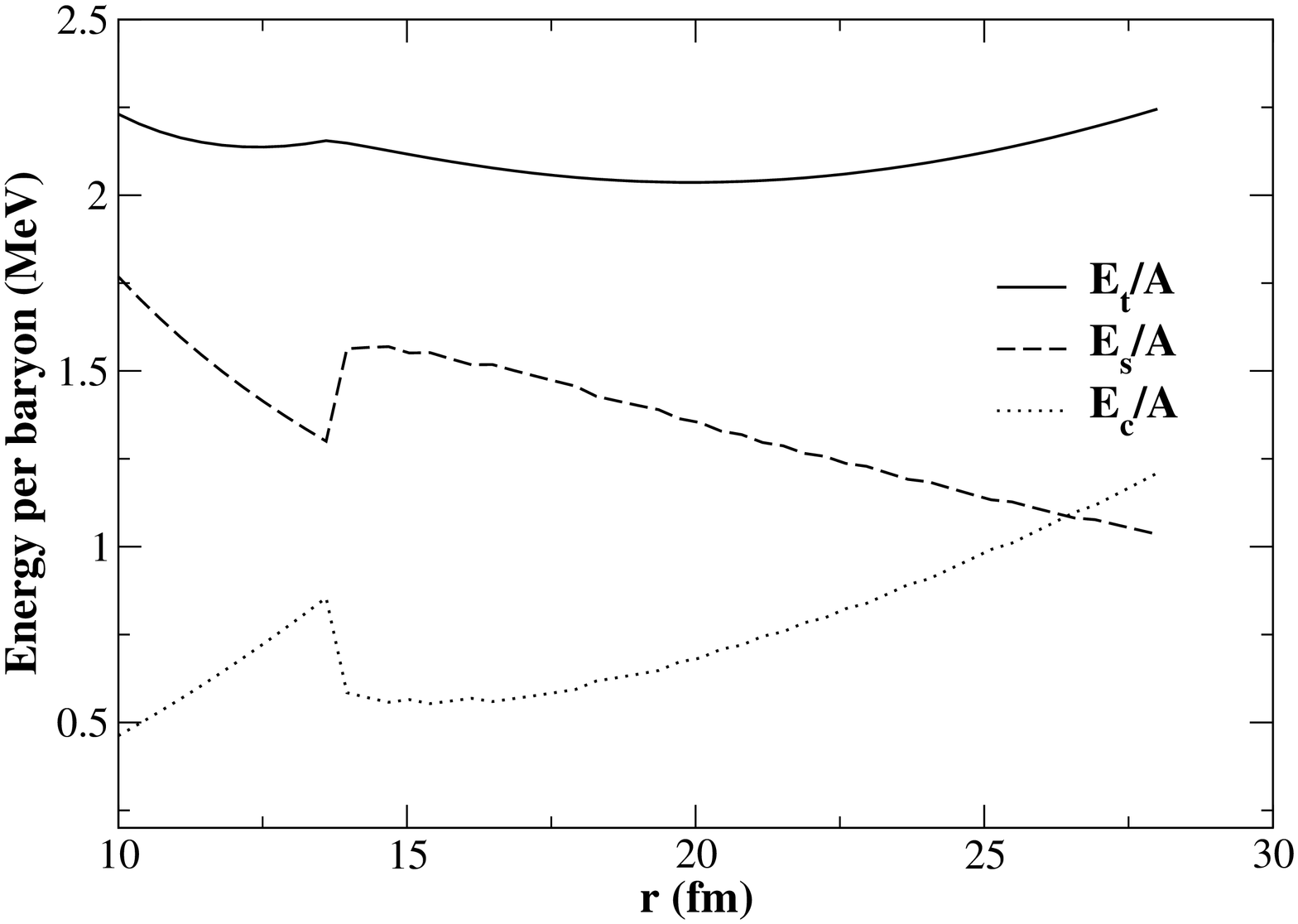}
 \caption{(Color online) Upper panel: nucleon density distributions for nucleus (solid black curve) and shell state (solid red curve)
 in a schematic model for WS cell with $\rho_B$ = 0.071 fm$^{-3}$ and $Y_p$ =
 0.5; dotted line is electron density. Lower panel:
the total energy per baryon $E_t/A$, surface energy per baryon
$E_s/A$ and Coulomb energy per baryon $E_c/A$ versus WS cell
radius, with the same conditions as above.}
 \label{toy}
\end{figure}

To get a more quantitative idea for the appearance of shell states,
we consider a simple model.  Suppose there is a spherical WS cell
with average baryon density $\rho_B$ = 0.071 fm$^{-3}$ and $Y_p$ =
0.5.  Assume the nucleons have a uniform density equal to the
saturation nuclear density.  There can be either a uniform density nucleus or a uniform density shell state with a hole in the center.  The schematic model is shown in the upper panel of Fig.~\ref{toy}.
The relevant contributions to the energy are the surface
and Coulomb energies.  For each fixed WS cell radius, the
radius of the outside, and perhaps inside surface of the nucleus can be adjusted to minimize the total energy.  By this procedure of minimization,
we can get the Coulomb energy, surface energy, and total energy of the cell for a range of cell radii as shown in the
lower panel of Fig.~\ref{toy}.  When the cell radius is small, the
configuration minimizing the total energy is that of a normal nucleus. When the cell radius increases, so that the Coulomb energy is larger
than half of the surface energy, the nucleons will
be pushed out from the center by Coulomb repulsion and form a shell.  This transition decreases the Coulomb energy while the surface energy is increased.  For large WS radii, the change in Coulomb energy dominates over the change in surface energy and the total energy of the cell will decrease.  Therefore the appearance of a shell state reduces the Coulomb energy and minimizes the total energy at the expense of a larger surface energy.  Now there are two minima in the total energy as a function  of WS cell radius. The nucleus state has a radius of 12.52 fm and has $^{A}Z$ =$^{584}292$. The shell state has a radius of 20.08 fm and $^{A}Z$= $^{2408}1204$.  Here the shell state is more stable since it has a lower energy.

%schematic model

There may be a similar situation for super heavy nuclei. For the
predicted doubly magic nucleus $^{292}120$, different mean field
model calculations show central depressions in the nuclear density
\cite{heavy}. The shell state we find can be considered as an even
larger super heavy nucleus with a central depression: the Coulomb
repulsion is so large that the protons are pushed out from the
center completely and locate in the outer region of the cell. Note
that the neutrons then follow the protons in order to reduce the
symmetry energy.

\begin{figure}[htbp]
 \centering
 \includegraphics[height=8cm,angle=-90]{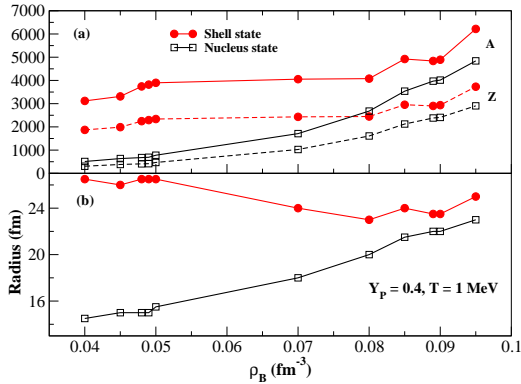}
 \caption{(Color online) The atomic and proton numbers (panel a),
 and the WS cell radii for nucleus (squares) and shell states (filled circles)
(panel b) shown as functions of the baryon density for $Y_p$ =
0.4.} \label{Yp0.4number}
\end{figure}

%Yp = 0.4, Rc vs density
In Fig.~\ref{Yp0.4number}, the atomic $A$ and proton $Z$ numbers
for nucleus and shell states (upper panel), and the corresponding
WS cell radii (lower panel) are shown as functions of the baryon
density.  The proton fraction is $Y_p$ = 0.4.  At low baryon
densities, the WS cell radius of the shell state is about twice
that of the nucleus state, {\it e.g.}, 26 fm vs. 14 fm at $\rho_B$
= 0.040 fm$^{-3}$. The cell radii of shell states decrease with
increasing density, while those of nucleus state increase with
density.  They approach one another at high baryon densities: 24
fm vs. 22 fm at 0.090 fm$^{-3}$.  For both nucleus and shell
states, $A$ and $Z$ increase with increasing baryon density.

%Yp = 0.4, transition to uniform matter

\begin{figure}[htbp]
 \centering
 \includegraphics[height=8cm,angle=-90]{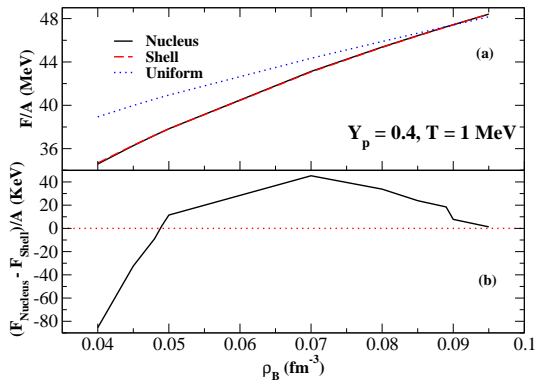}
 \caption{(Color online) Upper panel (a): the free energy per baryon for
non-uniform matter, nucleus (solid line) and shell states
(dot-dash line), and uniform matter (dashed line) are shown as
function of the baryon density for $Y_p$ = 0.4 and $T$ = 1 MeV.
Lower panel (b): free energy difference per baryon between nucleus
and shell states versus baryon density.} \label{Yp0.4_transition}
\end{figure}

In the upper panel of Fig.~\ref{Yp0.4_transition}, the free energy
per baryon for non-uniform matter (both nucleus state and shell
state) and uniform matter are shown as function of densities with
$Y_p$ = 0.4. In the lower panel, the free energy difference
between the nucleus and shell states is shown for various baryon
densities. This difference is of order tens of KeV and is maximum
around 0.070 fm$^{-3}$.  When the baryon density $\rho_B <$ 0.049
fm$^{-3}$, the nucleus state is the most stable.  Between
densities of 0.049 to 0.090 fm$^{-3}$, the shell state becomes
slightly more stable than the nucleus state.  When $\rho_B >$
0.090 fm$^{-3}$, the matter becomes uniform.

%Yp = 0.4: A and Z, WS radii vs baryon densities; F vs baryon densities.

% other Yps

\begin{figure}[htbp]
 \centering
 \includegraphics[height=8cm,angle=-90]{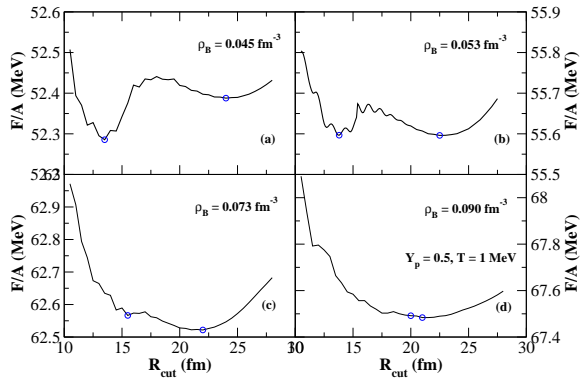}
 \caption{(Color online) Free energy per baryon versus Wigner-Seitz cell
 radius for baryon densities of $\rho_B$ = 0.045,
0.053, 0.073, and 0.090 fm$^{-3}$, proton fraction $Y_p = 0.5$ and
$T$ = 1 MeV. The two open circles denote minima corresponding to the nucleus (left) and shell (right) states.}
 \label{Yp0.5_mini}
\end{figure}

\begin{figure}[htbp]
 \centering
 \includegraphics[height=8cm,angle=-90]{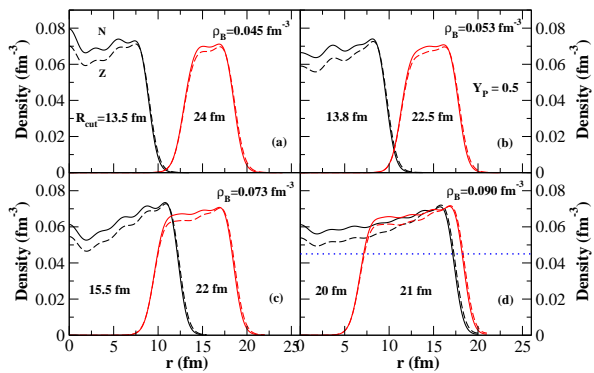}
 \caption{(Color online) The neutron (solid line) and
proton (dash line) density distributions for the Wigner-Seitz
cells with baryon density $\rho_B$ = 0.045, 0.053, 0.073, and
0.090 fm$^{-3}$, with proton fraction $Y_p = 0.5$ and $T$ = 1 MeV.
The red curves give density dist. of shell state. The black curves
give density dist. of nucleus state. The corresponding cell radii
are shown in each panel. The dotted line in the lower right panel
is the neutron or proton density of uniform matter.}
 \label{Yp0.5_density}
\end{figure}

\begin{figure}[htbp]
 \centering
 \includegraphics[height=8cm,angle=-90]{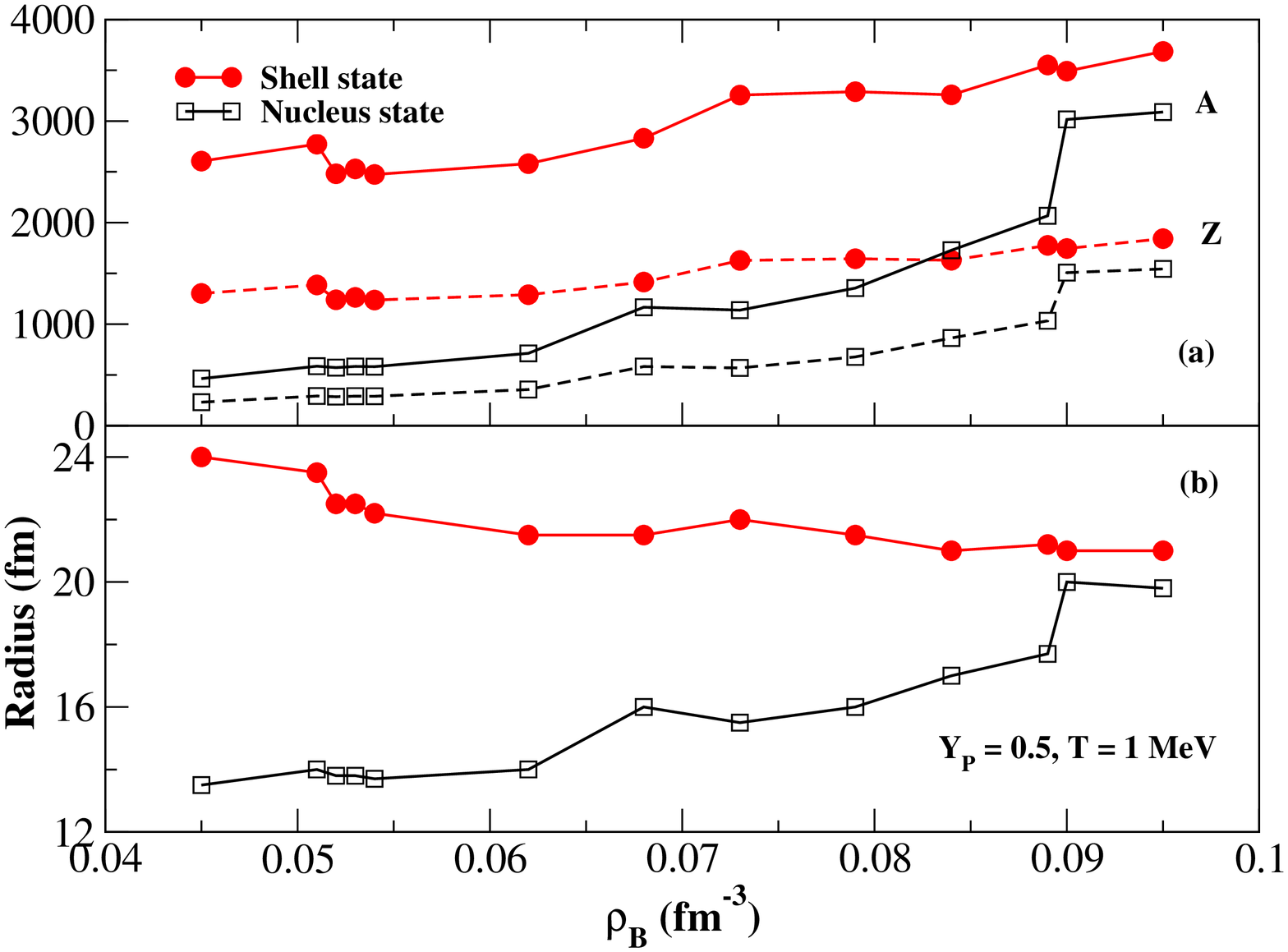}
 \caption{(Color online) The upper panel (a)
shows atomic (solid line) and proton (dashed line) numbers, while
the lower panel (b) shows WS cell radii, for nucleus (squares) and
shell states (filled circles) as functions of the baryon density
for $Y_p$ = 0.5.} \label{Yp0.5number}
\end{figure}

\begin{figure}[htbp]
 \centering
 \includegraphics[height=8cm,angle=-90]{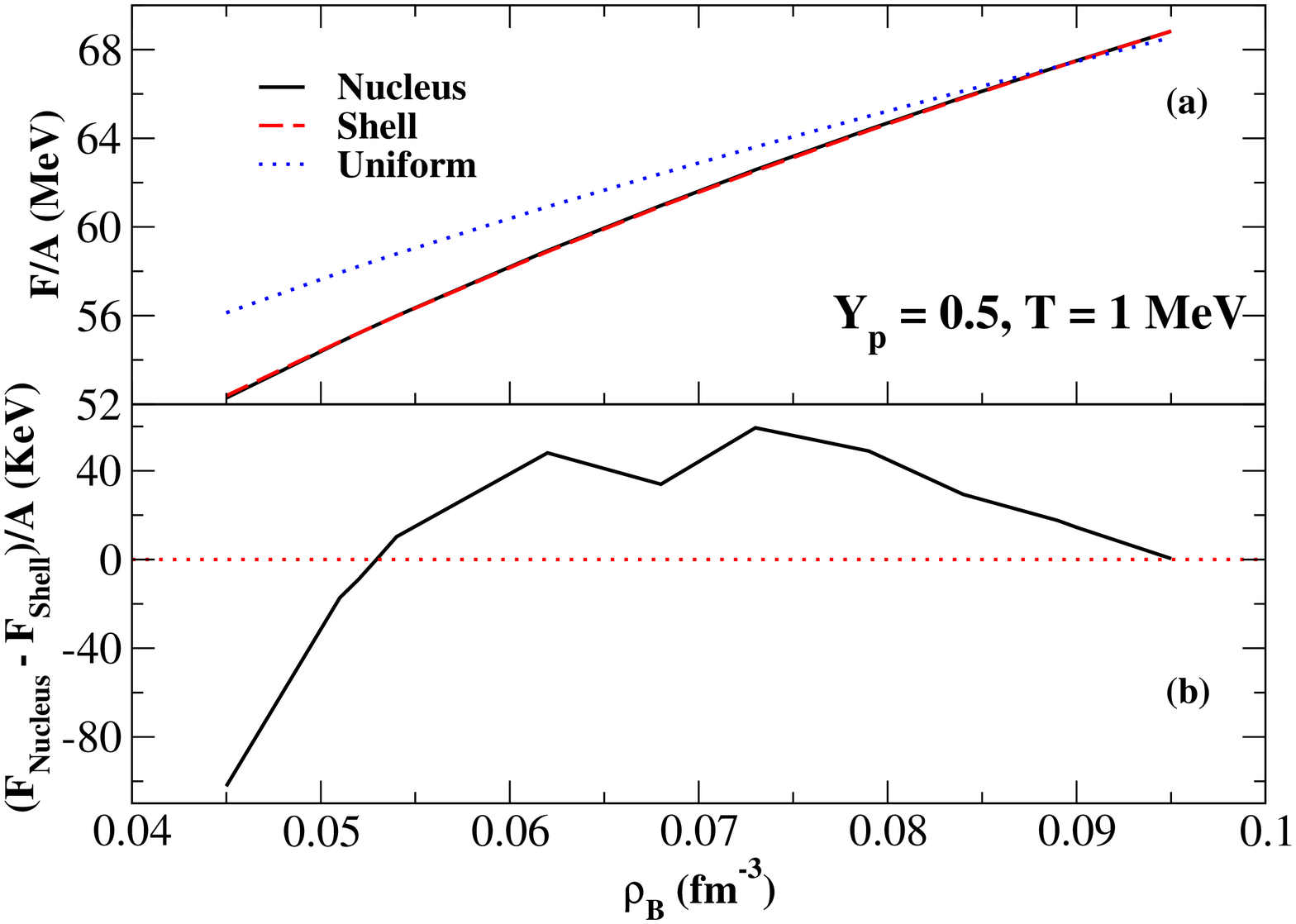}
 \caption{(Color online) Upper panel (a): the free energy per baryon for
non-uniform matter, nucleus (solid line) and shell state
(dot-dashed line), and uniform matter (dashed line) are shown as
functions of density for $Y_p$ = 0.5 and $T$ = 1 MeV. Lower panel
(b): free energy difference per baryon between nucleus and shell
states versus baryon density.} \label{Yp0.5_transition}
\end{figure}

The above analysis for $Y_p$ = 0.4 can be
generalized for different proton fractions $Y_p$.  For $Y_p$ = 0.5, the
minimization of free energy per baryon over the WS cell radii with
baryon density of 0.045, 0.053, 0.073, and 0.090 fm$^{-3}$ are
shown in Fig.~\ref{Yp0.5_mini}.  Again there are two minima
corresponding to nucleus state and shell state respectively. The neutron and proton density distributions associated with the two states are shown in Fig.~\ref{Yp0.5_density}. The transition density from nucleus
state to shell state is 0.053 fm$^{-3}$. In Fig.~\ref{Yp0.5number}, the atomic and proton numbers (upper panel), and the WS cell radii (lower panel) for nucleus state and shell state are shown for various baryon densities. In Fig.~\ref{Yp0.5_transition}, the free energy per baryon for
non-uniform matter (both nucleus state and shell state) and
uniform matter are shown as function of densities.  When the baryon
density is smaller than 0.053 fm$^{-3}$, the nucleus state is
the most favorable state. Between 0.053 to 0.090 fm$^{-3}$, the shell state become slightly more stable than the nucleus state. When the density is above 0.090 fm$^{-3}$ (same as that when $Y_p$ = 0.4), the uniform matter is the stable state.

\begin{figure}[htbp]
 \centering
 \includegraphics[height=8cm,angle=-90]{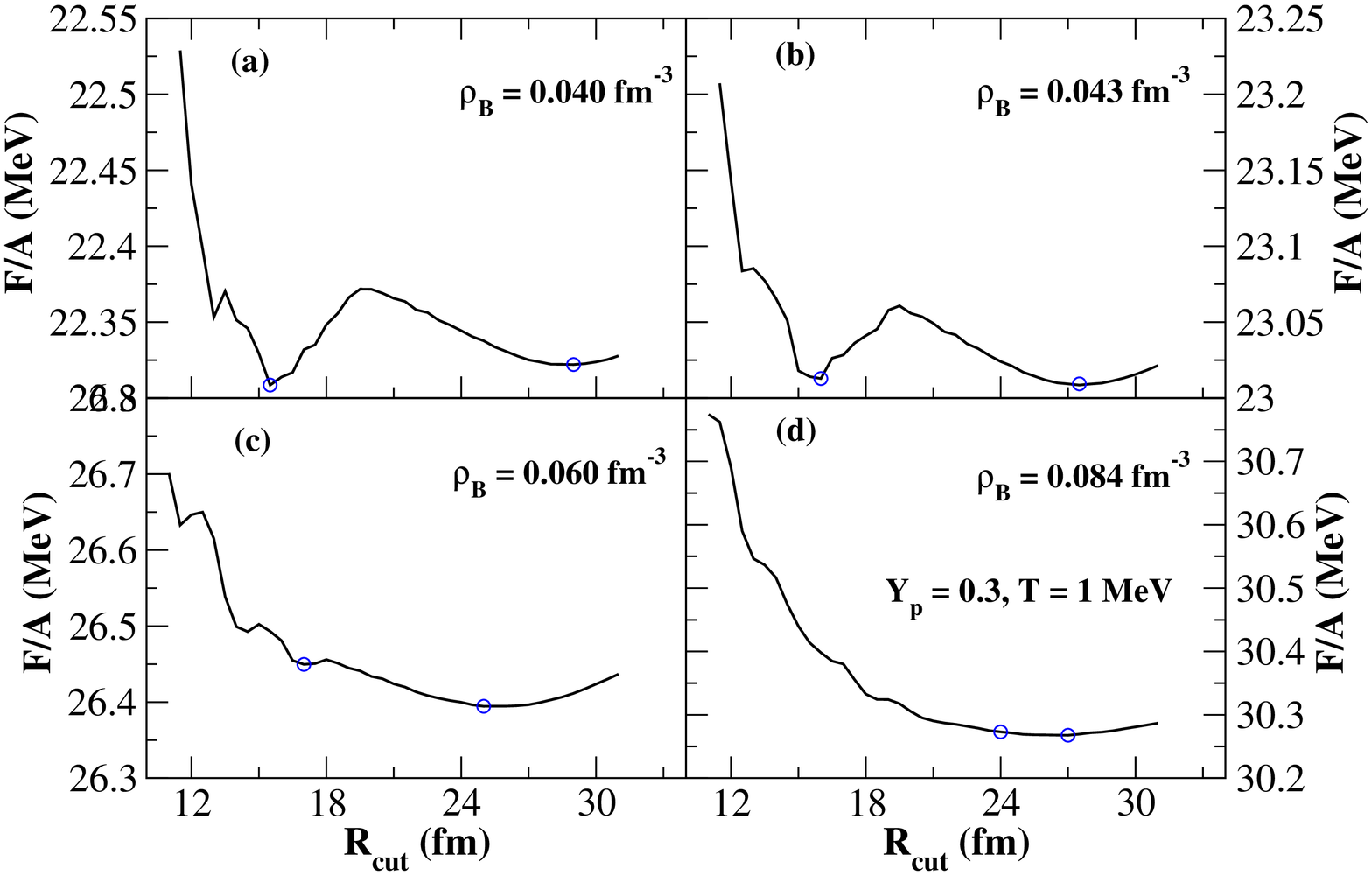}
 \caption{(Color online) Free energy per baryon versus Wigner-Seitz cell
 radii for baryon densities of $\rho_B$ = 0.040,
0.043, 0.060, and 0.084 fm$^{-3}$, proton fraction $Y_p = 0.3$ and
$T$ = 1 MeV. The two open circles denote minima corresponding to nucleus (left) and shell (right) states.}
 \label{Yp0.3_mini}
\end{figure}

\begin{figure}[htbp]
 \centering
 \includegraphics[height=8cm,angle=-90]{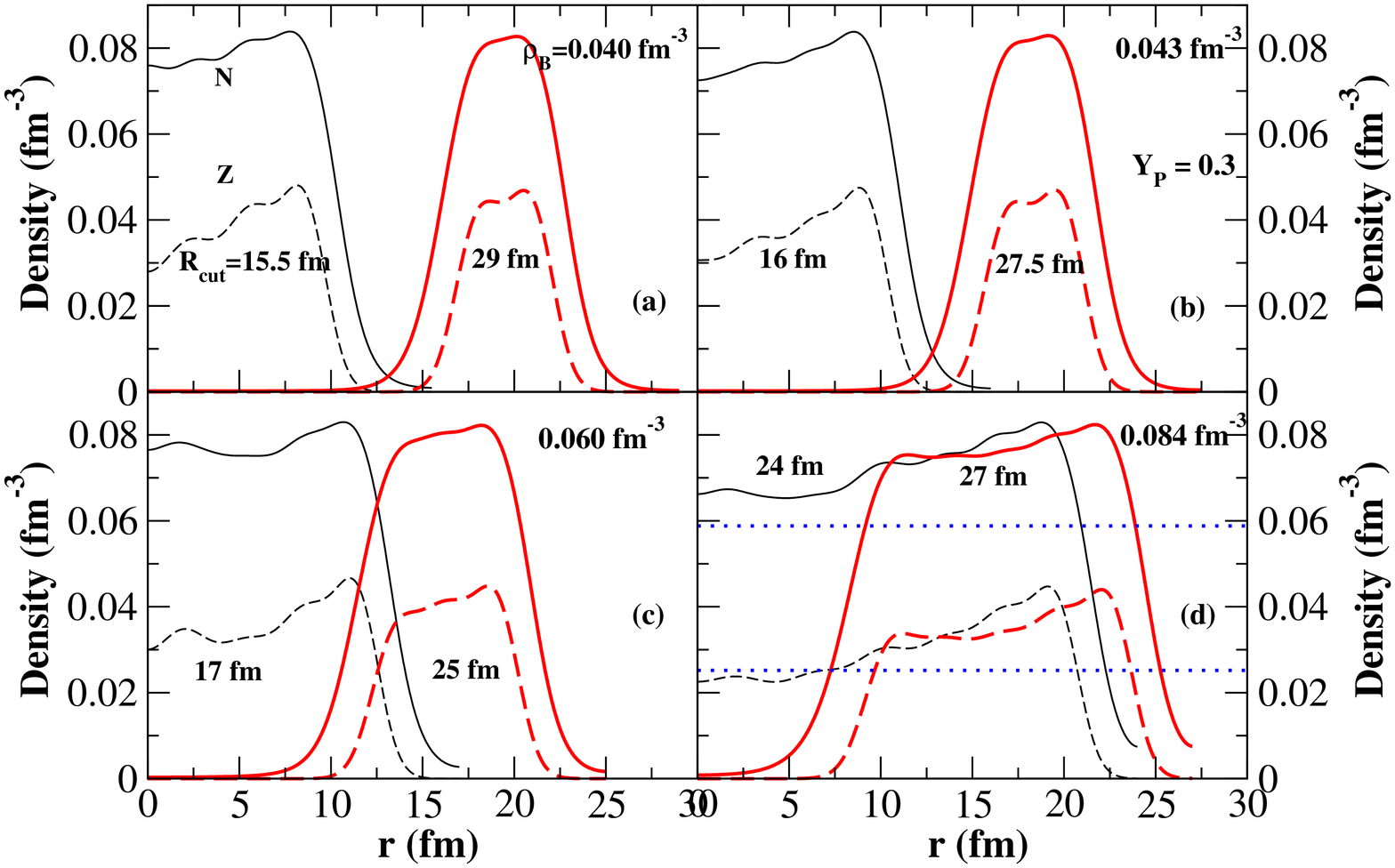}
 \caption{(Color online) The neutron (solid line) and
proton (dash line) density distributions for the Wigner-Seitz
cells at baryon densities of $\rho_B$ = 0.040, 0.043, 0.060, and
0.084 fm$^{-3}$, for proton fraction $Y_p = 0.3$ and $T$ = 1 MeV.
The red curves give density distributions for shell states. The
black curves give density distributions of nucleus states. The
corresponding cell radii are shown in each panel. The dotted lines
in the lower right panel are the neutron (upper) and proton
(lower) density of uniform matter.}
 \label{Yp0.3_density}
\end{figure}

\begin{figure}[htbp]
\bigskip
 \includegraphics[height=8cm,angle=-90]{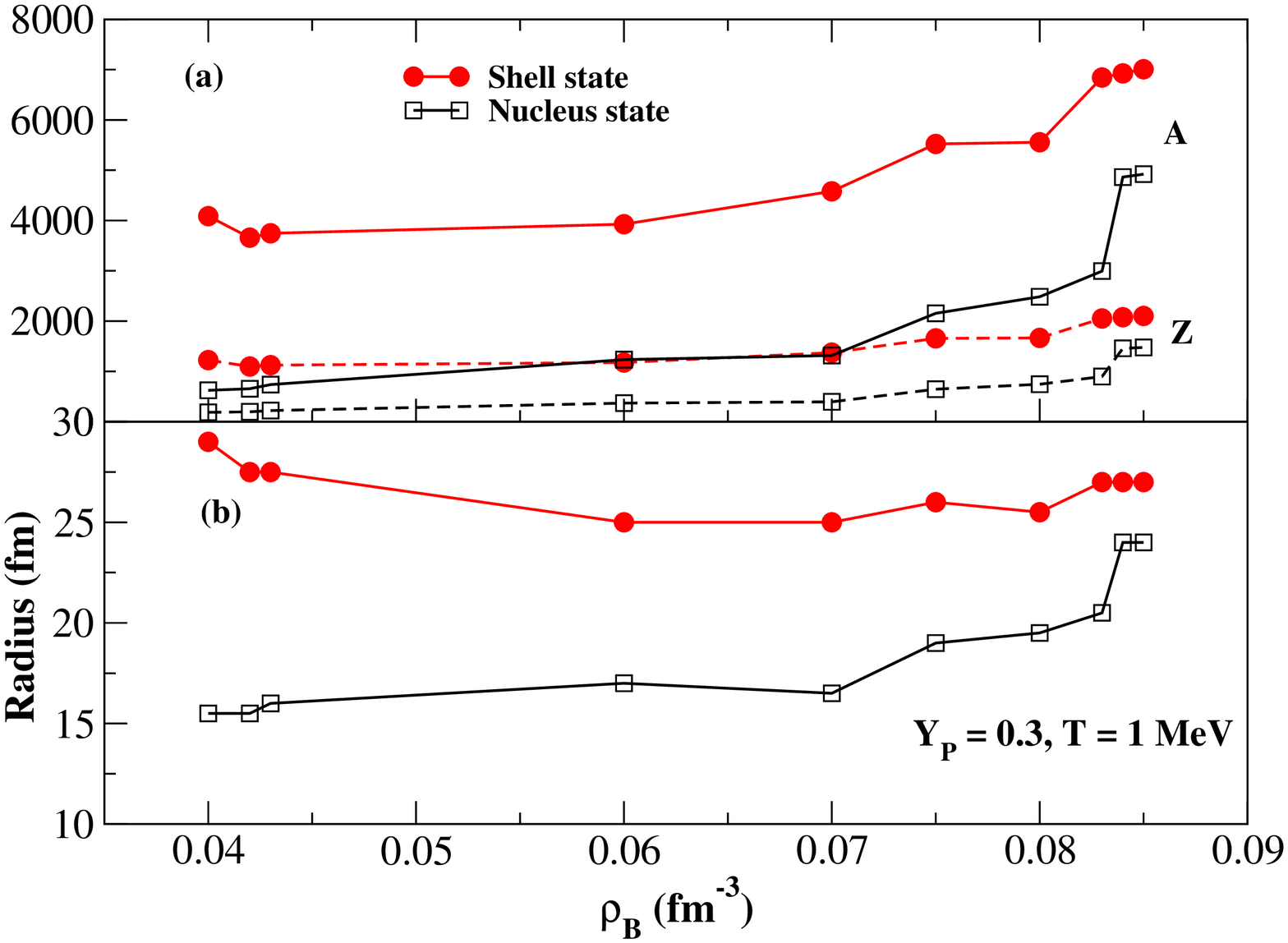}
 \caption{(Color online) The upper panel (a)
shows atomic (solid line) and proton (dashed line)  numbers, while
the lower panel (b) shows WS cell radii, for nucleus (squares) and
shell states (filled circles) as functions of the baryon density
for $Y_p$ = 0.3.} \label{Yp0.3number}
\end{figure}

\begin{figure}[htbp]
 \includegraphics[height=8cm,angle=-90]{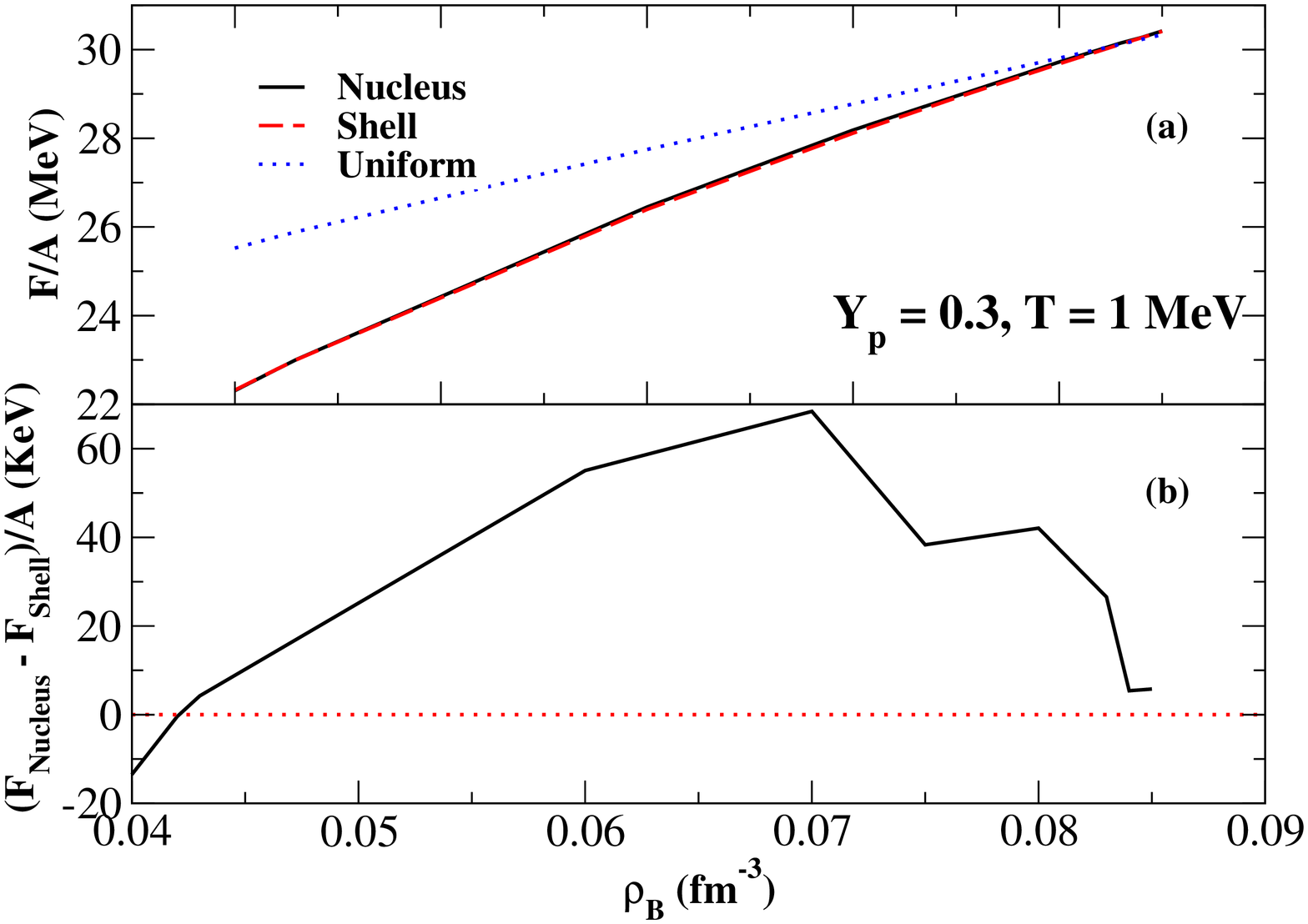}
 \caption{(Color online) Upper panel (a): the free energy per baryon for
non-uniform matter, nucleus (solid line) and shell state
(dot-dashed line), and uniform matter (dashed line) are shown as
functions of density for $Y_p$ = 0.3 and $T$ = 1 MeV. Lower panel
(b): free energy difference per baryon between nucleus and shell
states versus baryon density.} \label{Yp0.3_transition}
\end{figure}

\begin{figure}[htbp]
\includegraphics[height=8cm,angle=-90]{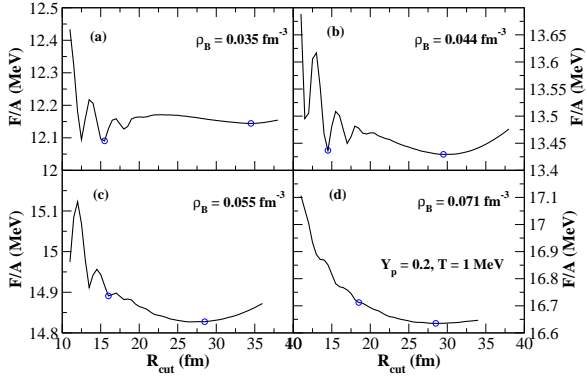}
 \caption{(Color online) Free energy per baryon versus Wigner-Seitz cell
radius for baryon densities of $\rho_B$ = 0.035, 0.044, 0.055, and
0.071 fm$^{-3}$, proton fraction $Y_p = 0.2$ and $T$ = 1 MeV. The
two open circles denote minima of nucleus (left) and shell (right)
states.}
 \label{Yp0.2_mini}
\end{figure}

\begin{figure}[htbp]
 \includegraphics[height=8cm,angle=-90]{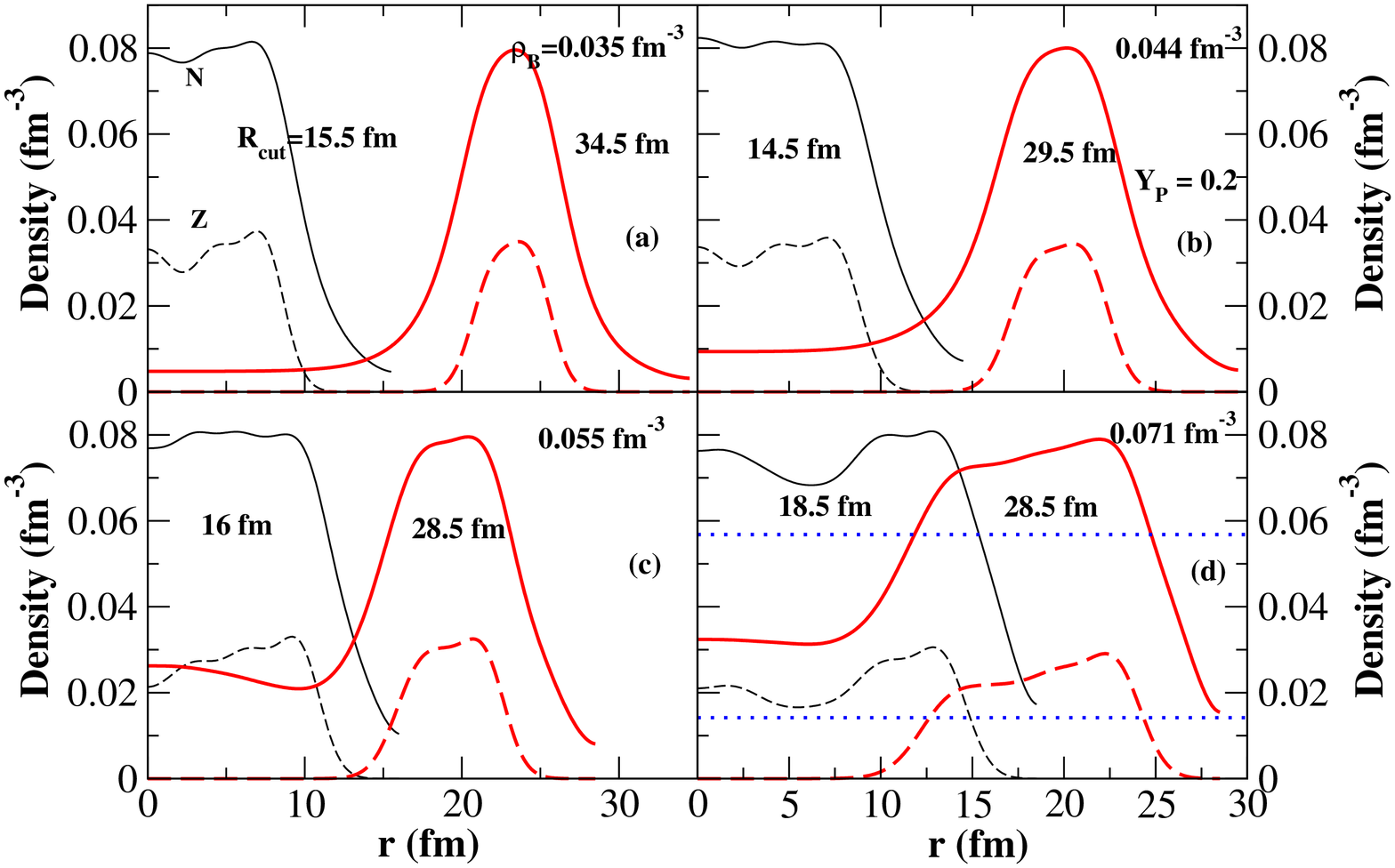}
 \caption{(Color online) The neutron (solid line) and
proton (dashed line) density distributions for the Wigner-Seitz
cells at baryon densities of $\rho_B$ = 0.035, 0.044, 0.055 and
0.071 fm$^{-3}$, with proton fraction $Y_p = 0.2$ and $T$ = 1 MeV.
The red curves give density distributions of shell states. The
black curves give density distributions of nuclear states. The
corresponding cell radii are shown in each panel. The dotted lines
in the lower right panel are the neutron (upper) and proton
(lower) densities of uniform matter.}
 \label{Yp0.2_density}
\end{figure}

\begin{figure}[htbp]
 \includegraphics[height=8cm,angle=-90]{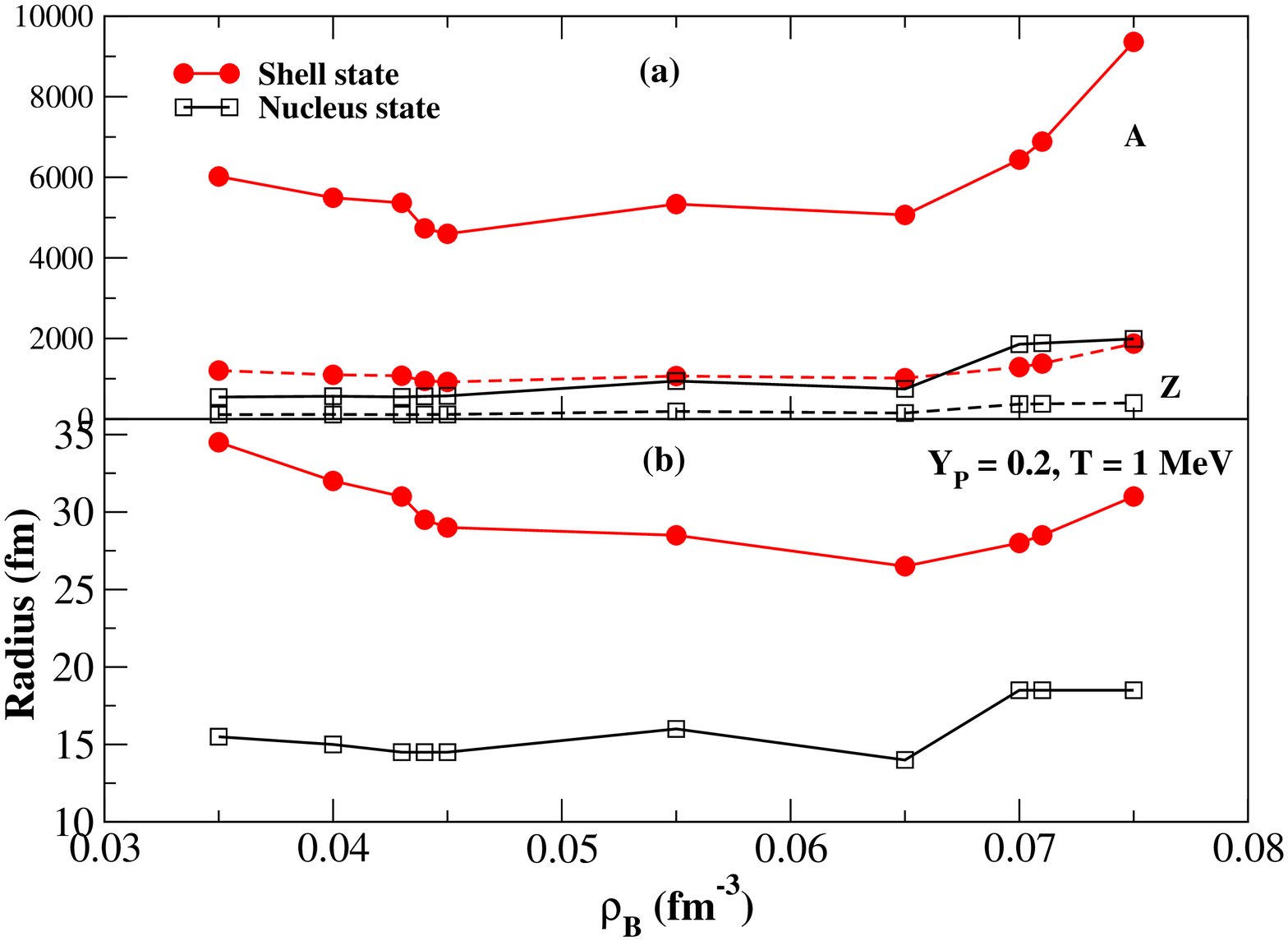}
 \caption{(Color online) The upper panel (a)
shows atomic (solid line) and proton (dashed line)  numbers, while
the lower panel (b) shows WS cell radii, for nucleus (squares) and
shell states (filled circles) as functions of the baryon density
for $Y_p$ = 0.2.} \label{Yp0.2number}
\end{figure}

\begin{figure}[htbp]
\includegraphics[height=8cm,angle=-90]{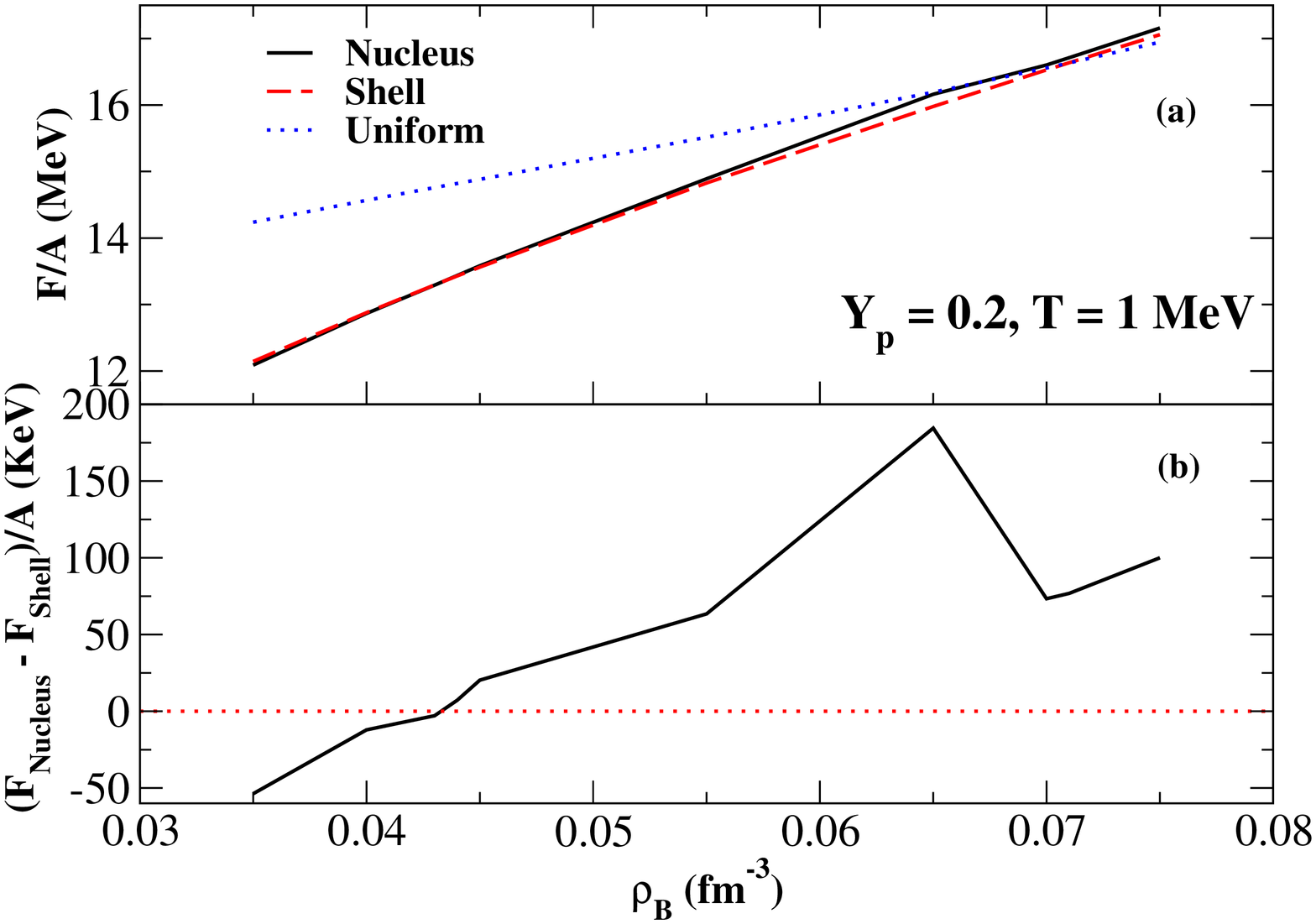}
 \caption{(Color online) Upper panel (a): the free energy per baryon for
non-uniform matter, nucleus (solid line) and shell state
(dot-dashed line), and uniform matter (dashed line) are shown as
function of density with $Y_p$ = 0.2 and $T$ = 1 MeV.  Lower panel
(b): free energy difference per baryon between the nucleus and
shell states versus baryon density.} \label{Yp0.2_transition}
\end{figure}

Similar calculations were carried out for $Y_p$ = 0.3 and the
results are presented in Figs.~\ref{Yp0.3_mini},
\ref{Yp0.3_density}, \ref{Yp0.3number} and \ref{Yp0.3_transition}.
We note that the transition density to the shell state is 0.043
fm$^{-3}$ and the transition density to uniform matter is 0.084
fm$^{-3}$. For $Y_p$ = 0.2, the results are presented in
Figs.~\ref{Yp0.2_mini}, \ref{Yp0.2_density}, \ref{Yp0.2number} and
\ref{Yp0.2_transition}.  The transition density to shell state is
0.044 fm$^{-3}$ and to uniform matter is 0.071 fm$^{-3}$. The
matter is very neutron rich at $Y_p$ = 0.2 and the nucleon density
distributions of shell states have a distinct feature, compared to
results for $Y_p$ = 0.3, 0.4, and 0.5. The neutron densities are
not vanishing in the center and on the edge of the shell nuclei,
as shown in Fig.~\ref{Yp0.2_density}. The neutron chemical
potential is greater than zero.  Therefore there are appreciable
numbers of free neutrons inside the cell.

% comparison of shell states for different yp

Comparison between Fig.~\ref{Yp0.5_transition} and
Fig.~\ref{Yp0.2_transition} indicates that shell states will start
to appear at smaller density for smaller $Y_p$.  Shell states involve a competition between surface and Coulomb energies.  Neutron rich matter, with only a few protons, has a
small binding energy and therefore a small surface energy. This
small surface energy makes it easier to form shell states.
Furthermore, we notice that at smaller $Y_p$ the transition to
uniform matter appears at smaller density. So in fact the density
range for the shell state has decreased for smaller $Y_p$.

% phase diagram at T = 1 MeV

\begin{figure}[htbp]
 \includegraphics[height=8cm,angle=-90]{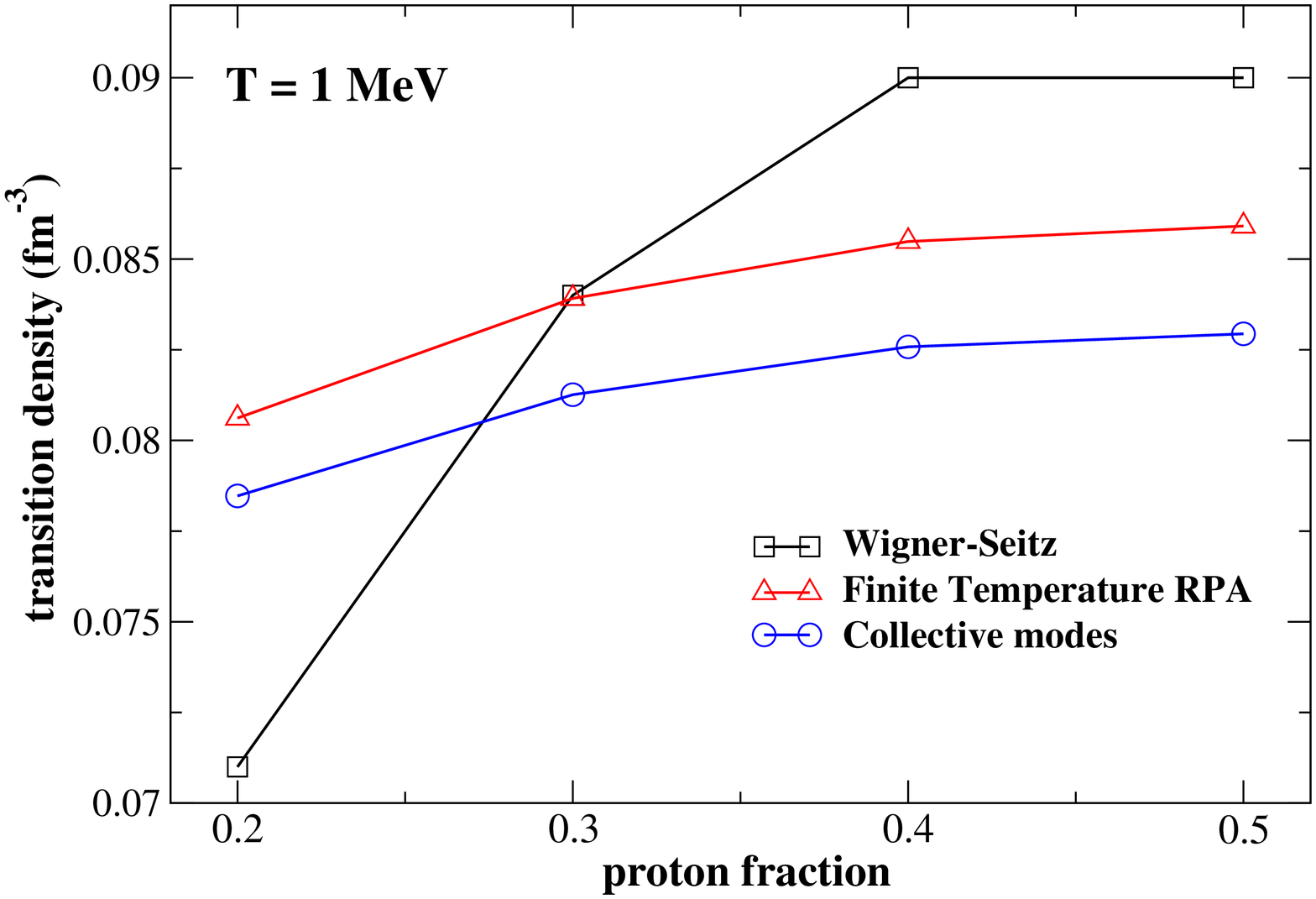}
 \caption{(Color online) The phase diagram for nuclear matter at $T$ = 1 MeV and different
proton fractions, combining results from Wigner-Seitz cell
calculation, collective modes analysis and RPA at finite
temperature.}
 \label{phase}
\end{figure}

Up to now, we have examined the EOS in the WS approximation for
nonuniform and uniform nuclear matter with $Y_p$ = 0.2, 0.3, 0.4,
and 0.5.  A universal picture of phase transitions from normal
nuclei to shell nuclei, then to uniform matter with increasing
baryon densities emerges.  It is beneficial to quantitatively
compare the results to other analytical methods.  As presented in
Section.~\ref{analysis}, using the eigencondition $D(\Omega=0)=0$ from
the semiclassical collective mode analysis and
Eq.~(\ref{ftcritic}) from the finite temperature RPA analysis
respectively, one can obtain the phase transition density to
uniform matter.  In Fig.~\ref{phase}, the phase transition
densities to uniform matter at $T$ = 1 MeV are shown for different
proton fractions. The transition densities obtained from the above two
analyses are plotted in comparison with the Hartree WS calculations.
At $Y_P$ = 0.3, these two analyses give the transition density
close to the WS approximation. When $Y_P
>$ 0.3, the two analyses give slightly smaller transition
densities. When $Y_P <$ 0.3, the two analyses give bigger
transition densities. The latter difference may suggest existence
of non-spherical configurations in the region between the spherical WS
transition density and the collective modes/RPA transition density for
non-uniform neutron rich matter.

%%%%%%%% Figures %%%%%%%%%%%%%%%%%%%%%%%%%

%Yp = 0.5: F vs Rc; density distributions.

%Yp = 0.5: A and Z, WS radii vs baryon densities; F vs baryon densities.

%Yp = 0.3: F vs Rc; density distributions.

%Yp = 0.3: A and Z, WS radii vs baryon densities; F vs baryon densities.

%Yp = 0.2: F vs Rc; density distributions.

%Yp = 0.2: A and Z, WS radii vs baryon densities; F vs baryon densities.

%phase diagram at T = 1 MeV

\section{\label{summary}Summary}

In this paper we applied a Relativistic Mean Field model to study the equation of state for non-uniform and uniform nuclear matter over a large range of baryon densities and with proton fractions, $Y_p$ = 0.2, 0.3, 0.4 and 0.5.  Within the spherical Wigner-Seitz approximation for non-uniform nuclear matter, we studied the system at a low temperature, $T$ = 1 MeV.  We find new shell states which minimize the free energy over a significant range of baryon densities.  These shell states minimize the  Coulomb energy of configurations with many protons at the expense of a larger surface energy.  These states are related to a possible central depression in the density distribution of super heavy nuclei.  As the density increases, the system undergoes phase transitions from normal nuclei to shell states, and then to uniform matter.

We compared our Wigner-Seitz results to two stability analyses of uniform nuclear matter.  Here we looked at collective modes in the RMF plus Vlasov formalism and in the finite temperature RPA. The
two analyses agree well with each other and give a slightly lower transition density to uniform matter at large proton fractions ($Y_p=0.5$, 0.4), and a higher transition density at small proton fractions (below 0.3) compared to the WS results.  This latter difference may suggest the existence of non-spherical configurations in the density region between the spherical WS transition density and the collective modes/RPA transition density for non-uniform neutron rich matter.

We conclude that even in a spherical approximation, there are more
possible configurations than have been previously considered.
These new shell states may play a role in the thermal and quantum
fluctuations of the system and in its response to a variety of
probes.  Furthermore these shell states may allow one to include,
in a simple way, much of the physics of, potentially complicated,
non-spherical pasta configurations.

In this investigation pairing between nucleons is not included,
since it is expected to have only a small effect on bulk properties such as the pressure or free energy.  However the effects of pairing on transport properties remains to be investigated.   In the future, we will study the equation of state at higher temperatures and investigate the response of shell states to supernova neutrinos.  We will also study the system at smaller proton fractions appropriate for neutron star matter in beta equilibrium.  Our ultimate goal is to produce an equation of state spanning a more complete range of baryon
densities, temperatures, and proton fractions, that will be
suitable for astrophysical applications to neutron stars and
supernovae.

\section{acknowledgement}
We thank Brian Serot for helpful discussions and Don Berry for
help with a computer cluster at Indiana University. This work was
supported in part by DOE grant DE-FG02-87ER40365.


\begin{thebibliography}{99}
\bibitem{heavyions} See for example C. Hartnack et al., Phys. Rev. Lett. {\bf 96}, 012302 (2006).  P. Danielewicz, R. Lacey, and W. G. Lynch, Science {\bf 22}, 1592 (2002).

\bibitem{x-rays} See for example S. Bogdanov et al., arXiv:0801.4030.  N. A. Webb and D. Barret, arXiv:0708.3816.

\bibitem{Negele} J.~W.~Negele and D.~Vautherin, Nucl. Phys. A {\bf
207}, 298 (1973).

\bibitem{more} See for example T. Maruyama et al., Phys. Rev. {\bf C72}, 015802 (2005).

\bibitem{crust1} D.~G.~Ravenhall, C.~J.~Pethick, and J.~R.~Wilson,
Phys. Rev. Lett. {\bf 50}, 2066 (1983).

\bibitem{crust2} C.~P.~Lorenz, D.~G.~Ravenhall, and C.~J.~Pethick, Phys. Rev.
Lett. {\bf 70}, 379 (1993).

\bibitem{sn} C.~J.~Horowitz, M.~A.~P\'{e}rez-Garc\'{i}a, and
J.~Piekarewicz, Phys. Rev. C {\bf 69}, 045804 (2204).


\bibitem{LSeos} J.~M.~Lattimer and F.~D.~Swesty, Nucl. Phys. A
{\bf 535}, 331 (1991).

\bibitem{HSeos} H.~Shen, H.~Toki, K.~Oyamatsu, and K.~Sumiyoshi,
Nucl. Phys. A {\bf 637}, 435 (1998).

\bibitem{HS1} C.~J.~Horowitz and B.~D.~Serot, Nucl. Phys. A {\bf 368}, 503 (1981).

\bibitem{HS2} C.~J.~Horowitz and B.~D.~Serot, Phys. Lett. B {\bf 86}, 146 (1979).

\bibitem{SW} B.~D.~Serot and J.~D.~Walecka, Adv. Nucl. Phys. {\bf{16}}, 1
(1986).

\bibitem{Rein}P.-G. Reinhard, Rep. Prog. Phys. {\bf {52}}, 439 (1989).
\bibitem{Ring}P.~Ring, Prog. Part. Nucl. Phys. {\bf {37}}, 193 (1996).

\bibitem{Kittel} C. Kittel, \textit{Introduction to Solid State Physics},
P. 571, John Wiley \& Sons Inc, 2nd Edition (1956).

\bibitem{Oyamatsu} K.~Oyamatsu, M.~Hashimoto and M.~Yamada, Prog. Theor. Phys. {\bf{72}}, 373 (1984).

\bibitem{Brito06} L.\ Brito, C.\ Providencia, A. M. Santos, S.\ S.\ Avancini, D.\ P.\ Menezes, and Ph.
Chomaz, Phys. Rev. C {\bf 74}, 045801 (2006).

\bibitem{ftRPA} C.~J.~Horowitz and K.~Wehrberger, Phys. Lett. B
{\bf 266}, 236 (1991).

\bibitem{heavy} A.~V.~Afanasjev, S.~Frauendorf, Phys. Rev. C {\bf 71}, 024308 (2005)
.

\bibitem{NL3}G.~A.~Lalazissis, J.~K\"{o}nig, and P.~Ring, Phys. Rev. C{\bf{55}}, 540 (1997).

\bibitem{Gogelein} See for example P. G\"{o}gelein, E. N. E. van Dalen, C. Fuchs, and H.
M\"{u}ther, Phys. Rev. C {\bf 77}, 025802 (2008).

\bibitem{pasta} H. Sonoda et al., Phys. Rev. {\bf C77}, 035806 (2008).  C. J. Horowitz et al., Phys. Rev. {\bf C70}, 065806 (2004).


\end{thebibliography}
\end{document}